**Dynamic manipulation of friction in smart textile composites of liquid-crystal elastomers**

*Takuya Ohzono, Mohand O. Saed, Youfeng Yue, Yasuo Norikane, Eugene M. Terentjev*

Dr. T. Ohzono, Dr. Y. Yue, Dr. Y. Norikane
National Institute for Advanced Industrial Science and Technology (AIST), 1-1-1 Higashi, Tsukuba 305-8565, Japan.    E-mail: ohzono-takuya@aist.go.jp

Dr. M. O. Saed, Prof. E. M. Terentjev
Cavendish Laboratory, University of Cambridge, J.J. Thomson Avenue, Cambridge, CB3 0HE, U.K.
E-mail: emt1000@cam.ac.uk

**Abstract:** Smart surfaces that reversibly change the interfacial friction coefficients in response to external stimuli enable a wide range of applications, such as grips, seals, brake pads, packaging films, and fabrics. Here a new concept of such a smart frictional system is reported: a composite film of a plain-weave polyester textile sheet, and a thermo-responsive nematic liquid crystalline elastomer (LCE). The composite is deployed with retractable micro-undulations of the elastomer inside each weave mesh, enabling dramatic changes of the contact interface with the opposing surface on LCE actuation, which is induced e.g. by a change in temperature ($T$). At room $T$, the protruding viscoelastic parts of LCE in the nematic phase make contact with the opposing flat surface, resulting in a very high friction. At an elevated $T$ (~50°C, isotropic phase), the undulations of LCE surface are retracted within the thickness of the textile, and the contacts are limited to small regions around overlapping textile fibers, lowering the friction dramatically. This effect is fully reversible on heating/cooling cycles. The surface undulations are spontaneous, i.e. fabricated without any lithographic or alignment techniques. The present composite opens a new way to practical uses of sheets/films with switchable friction enabled by stimuli-responsive LCEs.

---

Functional flexible polymer films and sheets[1] such as textiles, tapes, and laminating materials, are used in many applications. Their friction against other surfaces in contact[2–4] is one of the most important factors in their practical use, since this affects the force transmission between objects, e.g., in gripping and sliding, and in tactile sensing. In many situations, the ability to switch their friction characteristics on demand[5–9] would be useful. To design such smart contact surfaces, one or more of the parameters governing the friction should be altered through reversible changes in the material components[10,11].

The parameters controlling friction can be reviewed by considering a model for the dry friction in polymers and elastomers, which is known as the 'adhesive friction model'[2,3,12]. In this model, the friction force arises from the shear strength to open the contact interface. Including the energy dissipation effect due to the viscoelastic response, and the applied sliding speed, $v$, the overall friction force, $F$, may be written as,

$$F = \alpha F_0[1 + \phi(v)] \;, \tag{1}$$

where $F_0$ is the friction per unit area of real contacts at the zero crack opening speed, $\phi(v)$ is a non-dimensional factor expressing the velocity-dependent viscoelastic dissipation, and $\alpha$ is the ratio of the real contact area to the apparent one ($0<\alpha<1$). $F_0$ is related to the static interfacial energy of the two materials in contact. The geometric factor $\alpha$ should depend on the applied apparent normal pressure, $P$, (or load $L$), and usually shows a monotonic increase with $P$, known as the 'Amontons law'. Additionally, the variation in the elastic modulus of the materials in contact, which also affects the interfacial deformation, may be incorporated in $\alpha$. For the case of friction on the rough surface, the local topography effect can also be included in $\alpha$: the smaller or larger value should be assumed for the corrugated or smooth surface, respectively.

According to Eq. (1), there are, in principle, multiple parameters available to tune the friction. If materials in contact are changed, $F_0$ may change via the altered surface energies; p[arameters $\alpha$ and $\phi(v)$ also may change via the altered storage and loss moduli, respectively. If the surface topography is modulated, $\alpha$ is directly affected as well. Therefore, surface designs that induce drastic changes of materials in contact and the surface topography, will provide friction with a wide dynamic range.

For this purpose, ideas mainly relying on the surface topography changes have been explored, using mechanical buckling-based surface wrinkles[7–9,13] and fingerprint textures of a chiral nematic polymer coating[5,6,14]. The former system requires the application of macroscopic strain to the sample to generate the microscopic surface wrinkling, which makes it not useful as free-standing sheets. Although the latter liquid crystal polymer systems[15,16] are attractive, the good wrinkling effect requires specific alignment layers for the polymer on the surface, which is technically challenging and only useful in limited settings. Liquid crystal elastomers (LCEs) are very promising systems to induce very large topographical changes via phase transitions[17–25] that can be triggered by multiple stimuli, such as temperature, light, and chemicals. In addition, the LCEs show drastic changes in the bulk viscoelasticity[18,26–30], which allows to tune friction via the $F_0$, $\alpha$, and $\phi$ entries in the basic Eq. (1). However, the development of the dynamic frictional system through the combination of these effects remains a challenge.

Here, we propose a new composite system, combining a polyethylene (PE) textile and thermally responsive nematic LCEs. The generic idea is to, simultaneously: change the contact area through the topographic effects that influence the magnitude of $\alpha$, change materials in direct contact, which affects $F_0$ and $\phi$, and also use the dynamic soft-elasticity of a nematic LCE phase to separately affect $\phi$ via the enhanced internal dissipation. All of these changes occur on LCE actuation upon an external stimulus, e.g. temperature change that induces the phase transition of incorporated LCEs. Then, Eq. (1) is rewritten indicating the relevant variables as

$$F(T, L, v) = \alpha(T, L, v) F_0(T) [1 + \phi(T, v)] \ . \qquad (2)$$

The proposed changes become possible through a fine geometrical design of thermally retractable micro-undulations of LCE 'pockets' formed in each opening of the textile mesh. They are also designed to be fabricated through a simple lithography-free deposition method. At room temperature (RT), the buckled viscoelastic LCE parts in the nematic phase protrude out of the composite film and make contact with the opposing surface over a large area, resulting in high friction. In contrast, at high temperature (HT) that turns the LCE isotropic, the undulations of the elastomer are flattened and retracted within the confines of the textile mesh. As a result, the contacts with the opposing surface are limited to small regions around overlapping points of the hard textile fibers, lowering friction dramatically.

Our system is the composite (Figure 1) of a nematic LCE[23,25,30] and a plain-weave textile, in which the volume other than the fibre components within the thickness of the textile is almost fully filled with the LCE, as in the rubber-coated textiles. The basic properties of the present LCE have been characterized using the neat LCE samples. In brief, the LCE has $T_{NI}$ ~35°C, spontaneous and reversible changes of the shape with the uniaxial strain of at least 30%, and of the soft viscoelasticity with the high loss factor $tan\delta$ in the nematic phase[30].

The fabrication of the composite begins with molding the LCE precursor and the textile sheet between two flat glass slides under normal pressure[31], and its thermal curing in the isotropic phase (Fig. 1a). The first thiol-ene reaction, creates a random isotropic polymer network. During annealing of the peeled free-standing sample, the solvent is evaporated, causing the volumetric shrinkage. Since the textile lattice is locally rigid, the shrinkage occurs exclusively in the direction normal to the sample plane (Fig. 1b), which is different from the neat LCE systems without constraints of the textile frame.

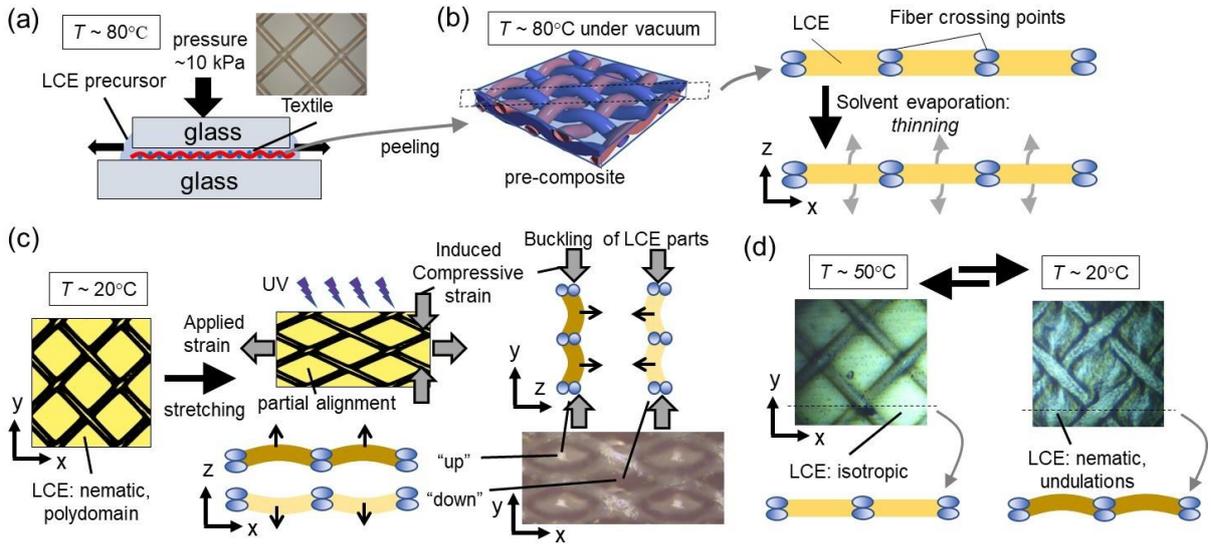

**Figure 1.** Preparation of the LCE-textile composite sheet with micro-shape-changeability: (a) The LCE precursor with the textile between two glass slides undergoes the first cross-linking reaction at elevated temperature (isotropic phase). (b) Drying the peeled sample under vacuum, prompting solvent evaporation and volume contraction. (c) The second network crosslinking by UV irradiation under local uniaxial strain at RT (nematic phase). The LCE parts buckle protruding in one of the out-of-plane directions, "up" or "down", alternatively along the nearest neighboring mesh openings. This structure is "memorized" by the secondary crosslinking in the nematic phase. (d) Release of applied tensile strain and annealing in isotropic phase causes restoring of the original shape, memorized in the isotropic state of LCE and the original textile shape with the square mesh opening. After cooling back to RT, the further modulated undulations of LCE segments appear due to the mismatch between the square shape of the mesh and the secondary memorized shape.

Then, the sample is uniaxially stretched at RT (in the nematic state) to the strain of 30% in the diagonal direction of the square mesh openings, which is the most compliant direction ($x$) of the plain-weave textile. Under this strain, the LCE parts at mesh openings adopt a buckled surface structure in response to the induced local compressive strain of approximately −45% in the orthogonal direction ($y$) (Fig. 1c).

Two energetically-equal states of buckling appear: bending "up" or "down", which is confirmed by the observation of both sides at the same location. In most parts on the sample, the buckling directions show a regular square pattern, with the nearest neigbouring mesh openings show ones opposite to each other. Since the LCE is continuous across neigbouring meshes, the alternation of the buckling direction may minimize the bending energy of LCE parts between the mesh openings. In addition to this structural change, the complicated alignment of the nematic director in this buckled state should emerge. This state of the LC alignment coupled to the shape is then memorized by the UV-induced second polymerization process that stabilizes the nematic phase.

After the second photo-crosslinking, the applied strain is released and the sample is annealed to equilibrate in a load-free state (Fig. 1d). At HT, the original memorized state of the isotropic LCE is restored, and the original textile shape with square mesh opening is also recovered. After cooling back to RT, the modulated undulations of LCE segments, reflecting those formed under the applied strain, appear without changing the square shapes of the mesh. Owing to the mismatch between the naturally square shape of the mesh openings, and the rhombic shape recorded in the nematic phase, additional buckling of the elastomer surface occurs in response to the emerging compressive strain in the $x$ direction. Since the initial undulations can assist in nucleating the additional buckling, the initial buckling directions memorized under applied strain are retained in the structure. Upon heating again to $T>T_{NI}$, the undulations retract, restoring the state of the first crosslinked network. Consequently, the *temperature* change induces reversible alternation of micro-undulations only, while the macroscopic shape of the composite sheet remains constant.

Since buckling of the LCE segments is the key phenomenon behind the micro-undulations, the effective aspect ratio, (lateral length)/(thickness), of the LCE segments in the mesh openings is an important parameter, which depends on the mesh size. In the subsequent study, the composites of the textile with the mesh opening of 0.199 mm (T100) are used as the representative system with ideal micro-undulations. The detailed temperature-dependent surface topographies are shown in Figure 2. The undulations of the LCE parts in the mesh openings at RT (nematic) are retracted at HT (isotropic) toward the mid-plane of the sample sheet. This is also traced by the diffusion of laser light transmited through the sample[32] showing the broadening of the diffusion peaks due to the roughening of the interface at RT.

The amount of the solvent during the preparation, $c_{sol}$, affects the surface topography via the final thickness of the LCE layer. In particular, it controls the height difference $h$ between the fiber-crossing points and the LCE parts (Fig. 2g, h). At HT, $h$ linearly increases at ~0.5 μm/wt% as shown by the red dashed line in Fig. 2h. Since the textile half thickness is ~50 μm, the result indicates that the shrinkage of LCEs near the center of the mesh opening occurs mainly in the direction perpendicular to the film. The non-zero value of $h$ at $c_{sol} = 0$ wt% would be due to the non-negligible curing shirinkage (syneresis)[33] of ~10% during the present polymerization. At RT, values of $h$ become smaller due to the micro-undulations at LCEs. The transformation between two states is fully repeatable over many cycles (Movie), as expected from the neat sample and the other nematic LCEs[17–25].

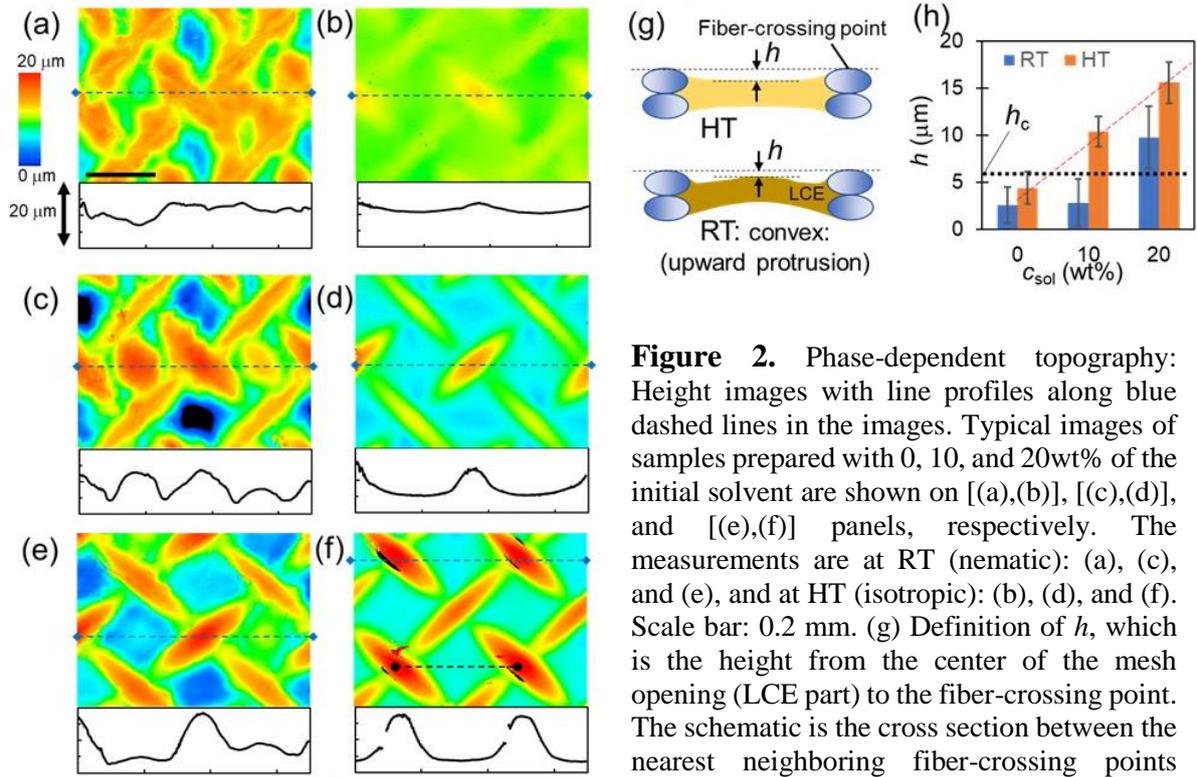

**Figure 2.** Phase-dependent topography: Height images with line profiles along blue dashed lines in the images. Typical images of samples prepared with 0, 10, and 20wt% of the initial solvent are shown on [(a),(b)], [(c),(d)], and [(e),(f)] panels, respectively. The measurements are at RT (nematic): (a), (c), and (e), and at HT (isotropic): (b), (d), and (f). Scale bar: 0.2 mm. (g) Definition of $h$, which is the height from the center of the mesh opening (LCE part) to the fiber-crossing point. The schematic is the cross section between the nearest neighboring fiber-crossing points indicated as dashed line in (f). For data at RT, the convex LCE parts were measured. (h) Plots of solvent concentration, $c_{sol}$, and averaged values of $h$ obtained from images with the larger area (Fig. S10) at multiple locations ($N > 10$). The characteristic height value, $h_c$ ~6 mm, dividing the two contact regimes described later, is shown as a guidance.

The static, $F_S$, and kinetic friction forces, $F_K$, between the LCE-textile composite and the flat glass surface were investigated (Figure 3a) at the range of load $L < 1$ N, for which the deformation of the PE fibre at the overlapping point is negligible. The typical time-dependent frictional signals are shown in Fig. 3b, in which $F_S$ and $F_K$ are defined. The temperature-dependent static and kinetic friction coefficients, $\mu_S$ and $\mu_K$, (Fig. 3c,d) are obtained as slopes of the linearly fitted line to the $L$-dependent friction (Fig. 3e,f). They are roughly classified as either the higher ($\mu_S > 1$, $\mu_K > 0.6$.) or lower ranges ($\mu_S < 0.4$, $\mu_K < 0.2$). According to the observation of contact regions, the high friction corresponds to the states with the larger contact area consisting of LCE parts (leading to a larger $\alpha$). In the low friction case, contacts only occur at points of overlapping fibres, with small area. In particular, the sample prepared with 10wt%-solvent shows maximum $T$-dependent changes in both $\mu_S$ and $\mu_K$, each of which shows almost six-fold increase on cooling from HT to RT. For comparison, the best results on the increase of friction in a LC polymer layer were at most 2-fold [6,14], and only when two patterned surfaces were against each other. In the deformation-induced wrinkled bilayer surface [7,8] the friction increase was at most 1.5-fold, see the review [13] for detail. The results demonstrate the effectiveness of the present composite design with $T$-dependent topographical changes on the dynamic friction.

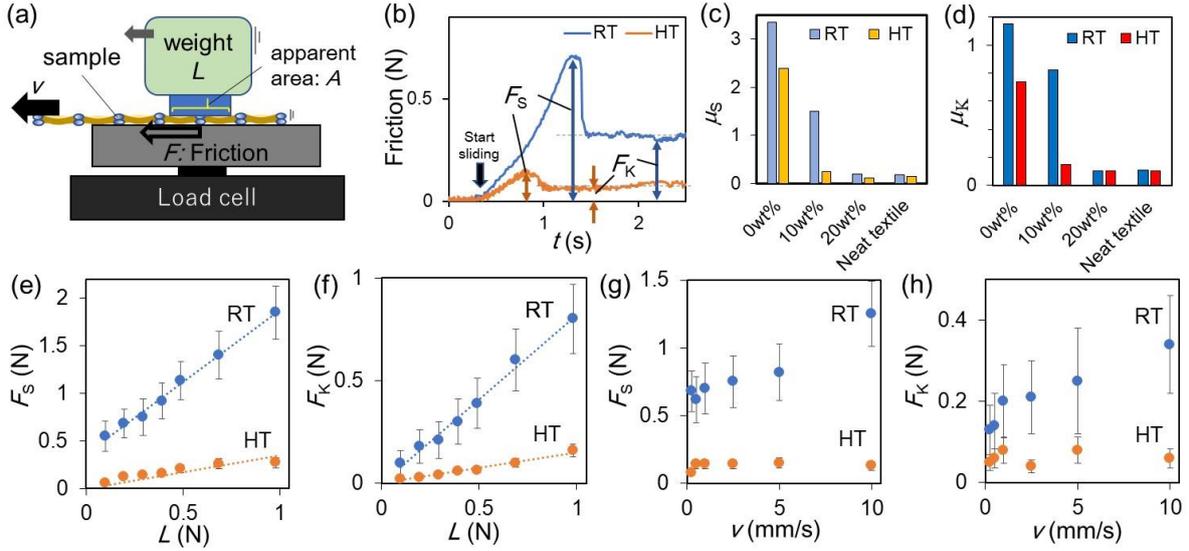

**Figure 3.** Friction tests: (a) Schematic for friction tests. (b) Example of the friction signal and the definition of static and kinetic frictions, $F_S$ and $F_K$, respectively. The data on the sample prepared with 10wt% solvent at $v$ = 5 mm/s with $L$ = 0.3 N (apparent normal pressure of 2.2 kPa) are shown. (c) Static, $\mu_S$, and (d) kinetic, $\mu_K$, friction coefficients, obtained as slopes of plots of friction force vs. normal load. (e)-(h) Plots of friction on composite prepared with solvent at 10wt%. Plots of (e) $F_S$, and (f) $F_K$, vs. normal load, $L$, at RT and HT. Plots of (g) $F_S$ and (h) $F_K$ vs. sliding speed $v$, at RT and HT.

Here let us briefly discuss the observed difference between $F_S$ and $F_K$. In general, $F_S$ at $L = 0$ may be in the same order of magnitude as the adhesive strength, $\sigma_{ad}$. Since $\sigma_{ad}$ between the glass surface and the LCE at RT after the dwell time of ~30 s is in the order of 0.1 MPa[30], $F_S$ at $L = 0$ with the apparent area of interface ($A$ = 133 mm$^2$) can be crudely estimated as $\alpha A \sigma_{ad}$ ~ 1 N, where $\alpha$~0.1 is the typical ratio of the contact area at $L = 0$ to $A$. This corresponds roughly to the non-zero values of the fitted line on the experiments at RT on 0wt%- and 10wt%-solvent systems at $L = 0$, giving ~0.9 N and ~0.4 N (Fig. 3e), respectively. The effect becomes smaller for the kinetic frictions (Fig. 3f), which show lines through the origin at $L = 0$, because of the interfacial sliding; no dwell time for the contact area growth.

Both $F_S$ and $F_K$ also increase with the sliding velocity, $v$ (Fig. 3g,h). The dependency is most striking at RT. The majority of the frictional interface is assumed to be the viscoelastic nematic LCE surface, with $tan\delta$ of ~0.7 or above, which should boost viscous dissipation, $\phi$, in Eq. (2). In contrast, in the isotropic phase at HT, the contact area is greatly reduced. Moreover, the LCE in contact is in the isotropic phase with the lower $tan\delta$ ~0.3 or below, and the hard PE surface at the fiber-crossing points are in contact, which supports most of the load. Note that at RT the friction of the sample prepared with 10wt%-solvent is only slightly larger than that observed on the neat textile (Fig. 3c,d). Thus the silght difference would originate from the modest contact with the isotropic LCE parts arround the fibre-crossing points. Overall, the present dynamic friction is qualitatively explained by $T$-dependent changes in $\alpha$ and $\phi$ in Eq. (2), which are related to the contact area and viscoelastic property, respectively.

In summary, here we report the design of LCE-textile composites with thermally retractable micro-undulations for the dynamic manipulation of friction coefficients via modulation of the contact states. The buckling-based spontaneous undulations of LCE parts in the openings of the textile mesh develop through a simple lithography-free procedure. The $T$-dependent interfacial contact states of the LCE-textile composites to a flat glass surface can be tuned by controlling the fine structure. With the optimized design, the LCE-textile composite makes contact at undulated viscoelastic LCE parts in the nematic phase at RT. In contrast, at HT, textile fibres at their crossing points mainly constitute the contacts with the small area. This alternation, which includes the changes in the contact area and replacement of the materials in contact, can modulate friction forces by sixfold. Owing to the simple fabrication of the transformable microstructure, and the possibility of the further optimization, including the fine structural modifications and tuning of LCE properties, the present design could provide a route to smart LCE-textile composite sheets and films that can reversibly change friction on demand. Since the textile[34] provides the LCE with a better mechanical strength, while retaining the flexibility of a free-standing film, the present composite would also be promising for handling and further processing in a wide range of applications, e.g., robot hands, touch screen devises, and smart textiles.

**Experimental Section**

*Materials and preparation of LCE-textile composite*: For preparation of LCE, we followed the methods reported previously,[25,30,35,36] with two steps of crosslinking reactions: a thiol-acrylate Michael addition, and a photoinduced radical polymerization of diacrylates, with slight modifications. The diacrylate monomer, 1,4-bis-[4-(6-acryloyloxyhexyloxy)benzoyloxy]-2-methylbenzene (RM82), was purchased from Wilshire Technologies. The diacrylate spacer, tri(propylene glycol) diacrylate (TPGDA), and two thiol monomers: 2,2'-(ethylenedioxy) diethanethiol (EDDET) and pentaerythritol tetrakis (3-mercaptopropionate) (PETMP), were purchased from Sigma Aldrich. Triethylamine (TEA, Sigma Aldrich) was used as the catalyst of the Michael-addition thiol-ene reaction. Irgacure2959 from BASF was used as the photoinitiator for the radial polymerization of acrylates. As the radical scavenger, butylated hydroxytoluene (BHT, from Sigma Aldrich) was used to supress the unwanted reaction before UV light irradiation. Toluene (Sigma Aldrich) was used as the solvent. All chemicals were used in their as-received condition with no purification. Three plain weave polyether (PE, polyethylene terephthalate) textiles composed of a fibre with the diameter of 0.055 mm and different mesh openings (0.368 mm (T60), 0.199 mm (T100), and 0.086 mm (T180)) were purchased from Yamani inc. Japan.

After the specific molar ratio of functional groups in RM82, TPGDA, EDDET and PETMP were weighed, Irgacure2959 was added at 0.2wt%, BHT was added at 0.5wt%, and toluene was added at 0, 10, and 20wt%. By increasing the amount of TPGDA, it was possible to lower $T_{NI}$ and here it was adjusted to make $T_{NI}$ ~35°C, that is, a temperature close to human hand. Note that the amount of the solvent was important to control the final thickness of LCE at the mesh openings, and thus varied here to study the effect. After the mixture was gently mixed at an elevated $T$ ~70°C for ~10 min, TEA was added at 1.5wt% to start the Michael-addition reaction between thiol and acrylate groups. The mixture was kept between two glass slides with a textile mesh with thickness of ~0.1 mm at 70°C (isotoropic phase) overnight under a pressure of ~10 kPa applied to the direction normal to the glass surface to squeese out the excess liquid[31]. The sample was released from the glass mold, and placed at 80°C in vacuum to remove the solvent. The sample was then cooled down to RT and UV light (365 nm) was applied for 20 min under a uniaxial tensile strain to finalize the crosslinking between the rest of the acrylates. The tensile strain of 30% was applied in the direction of 45° from the yarn direction; each mesh of the textile was thus elongated to a rhombic shape. The locally deformed and aligned state of the LCE segments were memorized at this crosslinking step. Upon releasing the strain, micro-undulations of the LCE segments were finally formed. For WAXS and mechanical measursments, samples without the mesh were also prepared, and UV light was applied under uniaxial tensile strain of 50% as reference. The samples were annealed at 80°C in vacuum oven for 12 hours before further characterizations.

*Characterization of surface topography*: The temperature-dependent surface topography was observed using the confocal laser reflection microscope (VK-9710, Keyence) with the temperature-controlled plate (Thermo Plate, Tokai-Hit).

*Observation of contact states*: The contact state was observed using inverted optical microscope (GX41, Olympus). The sample placed on a cover glass was set on the transparent hot plate (Thermo Plate, Tokai-Hit). From the top side of the sample, a flat glass surface (area of 133 mm$^2$) with a weight of 30 g (apparent pressure ~2.2 kPa) was gently placed. The contact regions, which appeared darker in most cases due to the change in the refractive index at the interface, were analyzed from the image.

*Friction tests*: The friction was evaluated under constant load and sliding speed using a setup, which was similar to the international-standard test method, ISO 8295 (Plastics: Film and sheeting: Determination of the coefficients of friction). First, the sample was placed on a glass surface connected to a load cell (Tribogear Type33, Heidon). A weight (10~100g) with a flat glass surface (area of $A = 133$ mm$^2$) was gently placed to apply constant normal load. After a dwell time of 30 s, during which the present LCE relaxed most of the strain[30], the sample with the weight was pulled (at a constant speed of $v$ of 0.25 to 10 mm/s) to start sliding between the sample surface and the bottom glass surface (Fig. 3a). The tests were done at ~20°C (RT) and ~50°C (HT) under the constant temperature control. The lateral force read by the load cell was measured as the time-dependent friction force. The static friction was defined as the peak value before the actual sliding at the interface of interest. The kinetic friction was obtained by averaging values during steady sliding over several seconds.

*Dynamic scanning callorimetry (DSC)*: For differential scanning calorimetry (DSC, DSC4000 PerkinElmer), samples with approximately 10 mg were loaded into standard aluminium DSC pans. The samples were heated to 90°C at 10°C/min, held isothermally for 5 min, and cooled to −60°C at 5°C/min to acquire the data. $T_{NI}$ can be found at local minimum of the endothermic peak. The sample was run three times.

*Dynamic mechanical analysis (DMA)*: The dynamic mechanical tests of neat LCE samples were performed on a Viscoanalyser-4000, Metravib, in the tension mode, with samples of 0.8 mm film thickness. A rectangular sample (effective length of 10 mm and width of 8 mm) was used. The simple strain of 0.5% was applied at frequencies of 5 and 50 Hz. Data were acquired on cooling at the rate of 3°C/min from 80 to −80°C.

*Stress-strain curves*: The stress-strain curves for the textiles, LCE-textile composites, and neat LCE films on the tensile mode were obtained using a commercial instrument (Tensilon EZ-LX, SHIMAZU). The sample width, thickness and effective length were, 5 mm, 0.09 mm and 30-50 mm, respectively. The uniaxial tensile stress-strain property in 45° from the axial direction of yarns, which is the softest direction, were evaluated. Separately, the curves in the uniaxial strain in 0° from the axial direction of yarns were obtain with T60 textile (a low density mesh). The strain rate of extension was 0.06-0.1% per second. Each measurement was run three times.

*Wide angle X-ray scattering (WAXS)*: The phase of the present LCE at RT was characterized using a Philips diffractometer using a Philips Copper target (PW-2233/20) with the wavelength of 0.154 nm. The distance between the sample and the imaging area was 100 mm.

*Fourier Transform Infrared Spectroscopy (FTIR)*: The FTIR measurements were performed using a Nicolet iS10 FTIR spectrometer, Thermo scientific. The peaks of thiols and acrylate at 2573 cm$^{-1}$ and 811 cm$^{-1}$, respectively, were evaluated. Samples were prepared and mounted between KBr plates. The initial state was a sample without TEA. The sample placed under 70°C oven for 24 hours after addition of TEA was also prepared. After taking the spectrum, UV light was irradiated for 20 min for evaluation of the final state. The conversion ratios at the end of each reaction were defined as $\xi = \frac{\sigma_{\text{final}} - \sigma_{\text{initial}}}{\sigma_{\text{initial}}}$, where $\sigma_{\text{initial}}$ and $\sigma_{\text{final}}$ are the initial an final peak area of each functional group, thiol or acrylate.

**References**

[1] S. Schneegass, O. Amft, *Smart Textiles Fundamentals, Design, and Interaction*, Springer, Cham, Switzerland, **2017**.
[2] D. F. Moore, *The Friction and Lubrication of Elastomers*, Pergamon, Oxford, **1972**.
[3] B. N. J. Persson, *Sliding Friction Physical Principles and Applications*, Springer, Heidelberg, **2000**.
[4] V. L. Popov, *Contact Mechanics and Friction Physical Principles and Applications*, Springer, Heidelberg, **2010**.
[5] D. Liu, D. J. Broer, *Soft Matter* **2014**, *10*, 7952.
[6] D. Liu, D. J. Broer, *Angew. Chemie - Int. Ed.* **2014**, *126*, 4630.
[7] K. Suzuki, Y. Hirai, T. Ohzono, *ACS Appl. Mater. Interfaces* **2014**, *6*, 10121.
[8] K. Suzuki, Y. Hirai, M. Shimomura, T. Ohzono, *Tribol. Lett.* **2015**, *60*, 2.
[9] K. Suzuki, T. Ohzono, *Soft Matter* **2016**, *12*, 6176.
[10] J. Hu, H. Meng, G. Li, S. I. Ibekwe, *Smart Mater. Struct.* **2012**, *21*, 053001.
[11] M. Wei, Y. Gao, X. Li, M. J. Serpe, *Polym. Chem.* **2017**, *8*, 127.
[12] K. Grosch, *Nature* **1963**, *197*, 858.
[13] N. Hu, R. Burgueño, *Smart Mater. Struct.* **2015**, *24*, 063001.

[14] D. Liu, D. J. Broer, *Langmuir* **2014**, *30*, 13499.
[15] Y. Yu, M. Nakano, T. Ikeda, *Nature* **2003**, *425*, 145.
[16] Y. Yue, Y. Norikane, R. Azumi, E. Koyama, *Nat. Commun.* **2018**, *9*, 3234.
[17] J. Küpfer, H. Finkelmann, *Mocromolecular Rapid Commun.* **1991**, *12*, 717.
[18] E. M. Warner, M., Terentjev, *Liquid Crystal Elastomers*, Oxford Univ. Press, **2007**.
[19] A. Sanchez, H. Finkelmann, *Solid State Sci.* **2010**, *12*, 1849.
[20] T. J. White, D. J. Broer, *Nat. Mater.* **2015**, *14*, 1087.
[21] V. P. Tondiglia, T. J. White, *Science* **2015**, *347*, 982.
[22] A. D. Auguste, J. W. Ward, J. O. Hardin, B. A. Kowalski, T. C. Guin, J. D. Berrigan, T. J. White, *Adv. Mater.* **2018**, *30*, 1802438.
[23] M. O. Saed, C. P. Ambulo, H. Kim, R. De, V. Raval, K. Searles, D. A. Siddiqui, J. M. O. Cue, M. C. Stefan, M. R. Shankar, T. H. Ware, *Adv. Funct. Mater.* **2019**, *29*, 1806412.
[24] M. Barnes, R. Verduzco, *Soft Matter* **2018**, *15*, 870.
[25] C. M. Yakacki, M. Saed, D. P. Nair, T. Gong, S. M. Reed, C. N. Bowman, *RSC Adv.* **2015**, *5*, 18997.
[26] S. M. Clarke, A. R. Tajbakhsh, E. M. Terentjev, C. Remillat, G. R. Tomlinson, J. R. House, *J. Appl. Phys.* **2001**, *89*, 6530.
[27] E. M. Terentjev, A. Hotta, S. M. Clarke, M. Warner, G. Marrucci, J. M. Seddon, H. S. Sellers, *Philos. Trans. R. Soc. A Math. Phys. Eng. Sci.* **2003**, *361*, 653.
[28] S. M. Clarke, A. R. Tajbakhsh, E. M. Terentjev, M. Warner, *Phys. Rev. Lett.* **2001**, *86*, 4044.
[29] D. R. Merkel, R. K. Shaha, C. M. Yakacki, C. P. Frick, *Polymer* **2019**, *166*, 148.
[30] T. Ohzono, M. O. Saed, E. M. Terentjev, *Adv. Mater.* **2019**, *31*, 1902642.
[31] T. Ohzono, K. Teraoka, *R. Soc. open sci.* **2018**, *5*, 181169.
[32] T. Ohzono, K. Suzuki, T. Yamaguchi, N. Fukuda, *Adv. Opt. Mater.* **2013**, *1*, 374.
[33] S. Zhiquan, *J. Mater. Chem. C* **2018**, *6*, 11561.
[34] D. Rosato, D. Rosato, *Reinforced Plastics Handbook*, Elsevier Science, Oxford, **2004**.
[35] D. P. Nair, N. B. Cramer, J. C. Gaipa, M. K. McBride, E. M. Matherly, R. R. McLeod, R. Shandas, C. N. Bowman, *Adv. Funct. Mater.* **2012**, *22*, 1502.
[36] M. O. Saed, R. H. Volpe, N. A. Traugutt, R. Visvanathan, N. A. Clark, C. M. Yakacki, *Soft Matter* **2017**, *13*, 7537.